\documentstyle[11pt]{article}
\begin{document}
\title{\Large\bf  Strange axial-vector mesons mixing angle }

\author{\small De-Min Li\footnote{E-mail: lidm@zzu.edu.cn},~~Zhen Li\\
\small  Department of Physics, Zhengzhou University, Zhengzhou, Henan 450052, P. R. China\\}
\date{\today}
\maketitle \vspace{0.5cm}

\begin{abstract}

The masses of the $K_1(^3P_1)$ and $K_1(^1P_1)$ are considered in a nonrelativistic constituent
quark model, and  the absolute value of the $K_1(^3P_1)-K_1(^1P_1)$ mixing angle is determined to
be about $59.29^\circ$. Comparison of the theoretical predictions on the strong decay widths of the
$K_1(1270)$ and $K_1(1400)$ in the $^3P_0$ decay model as well as the production ratio of these two
states in the $\tau$ decay between the available experimental data strongly favors that the
$K_1(^3P_1)-K_1(^1P_1)$ mixing angle is about $+59.29^\circ$.
\end{abstract}

\vspace{0.5cm}


Decays of taus


\newpage

\baselineskip 24pt

\section{Introduction}
\indent \vspace*{-1cm}

 The strange axial vector mesons provide interesting possibilities to study the QCD in the nonperturbative regime by
 the mixing of the $^3P_1$ and $^1P_1$ states. In the exact SU(3) limit, the $K_1(^3P_1)$ and
 $K_1(^1P_1)$ do not mix, just as the $a_1$ and $b_1$ mesons do not mix.
 For the strange quark mass
 greater than the up and down quark masses so that SU(3) is broken, also, the $K_1(^3P_1)$ and
 $K_1(^1P_1)$ do not possess definite C-parity, therefore these states can in principle mix
 to give the physical $K_1(1270)$ and $K_1(1400)$.

 Accurate determination of $\theta_K$, the mixing angle of the $K_1(^3P_1)$ and
 $K_1(^1P_1)$, is important for comparing the theory predictions about the decays involving the strange
 axial-mesons with the experimental data. In the literature,
 $\theta_K$ has been estimated by some different approaches, however, there is not yet a
 consensus on the value of $\theta_K$.  As the optimum fit to the data as of
 1977, Carnegie et al. finds $\theta_K=(41\pm 4)^\circ$\cite{carnegie}. Within the heavy quark effective theory
Isgur and Wise predict two possible mixing angles, $\theta_K\sim 35.3^\circ$ and $\theta_K\sim
-54.7^\circ$\cite{isgur}. Based on the analysis of $\tau\rightarrow \nu K_1(1270))$ and
$\tau\rightarrow \nu K_1(1400))$, Rosner suggests $\theta_K\sim 62^\circ$\cite{rosner},  Asner et
al. gives $\theta_K= (69\pm 16\pm 19)^\circ$ or $(49\pm 16\pm 19)^\circ$\cite{asner}, and Cheng
obtains $\theta_K= \pm 37^\circ$ or $\pm 58^\circ$\cite{cheng}. From the experimental information
on masses and the partial rates of $K_1(1270)$ and $K_1(1400)$, Suzuki finds two possible solutions
with a two-fold ambiguity, $\theta_K\sim 33^\circ$ or $57^\circ$\cite{suzuki}. A constraint
$35^\circ\leq \theta_K \leq 55^\circ$ is predicted by Burakovsky et al. in a nonrelativistic
constituent quark model\cite{burakovksy}, and within the same model, the values of $\theta_K\simeq
(31\pm 4)^\circ$ and $\theta_K\simeq (37.3\pm 3.2)^\circ$ are suggested by Chliapnikov\cite{nrqm}
and Burakovsky\cite{prd57}, respectively. The calculations for the strong decays of $K_1(1270)$ and
$K_1(1400)$ in the $^3P_0$ decay model suggest $\theta_K\sim 45^\circ$\cite{blundell,barnes}. The
mixing angles $\theta_K\sim 34^\circ$\cite{isgod}, $\theta_K\sim 5^\circ$\cite{godkok} are also
presented within a relativized quark model. Vijande et al. suggests $\theta_K\sim 55.7^\circ$ based
on the calculations in a constituent quark model\cite{vijande}. More recently, based on the
$f_1(1285)-f_1(1420)$ mixing angle $\sim 50^\circ$ derived from the analysis for a substantial body
of data concerning the $f_1(1420)$ and $f_1(1285)$\cite{zphysc76}, we suggest that the
$K_1(^3P_1)-K_1(^1P_1)$ mixing angle is about $\pm (59.55\pm 2.81)^\circ$\cite{epja26}.

In the present work, we shall show that the $K_1(^3P_1)-K_1(^1P_1)$ mixing angle derived from the
nonrelativistic constituent quark model is in good agreement with that given by Ref.\cite{epja26},
and try to constrain the sign of the $K_1(^3P_1)-K_1(^1P_1)$ mixing angle by considering the
open-flavor strong decays of the $K_1(1270)$ and $K_1(1400)$ in the $^3P_0$ decay model and the
production ratio of these two states in the $\tau$ decay.

\section{Nonrelativistic constituent quark model for $P$-wave mesons}
 \indent\vspace*{-1cm}

In the constituent quark model, the conventional $q\bar{q}$ wave function is typically assumed to
be a solution of a nonrelativistic Schr$\ddot{\mbox{o}}$dinger equation with the generalized
Breit-Fermi Hamiltonian which contains a QCD inspired potential $V(\mbox{\bf r})$\cite{qm1}. The
phenomenological forms of the matrix element of the Breit-Fermi Hamiltonian for the $q\bar{q}$
mesons with orbital angular momentum $L$ are given by\cite{nrqm,qm2}:
\begin{eqnarray}
M_{L=0}&=&m_q+m_{\bar{q}}+e_0\frac{\langle\mbox{\bf s}_q\cdot\mbox{\bf s}_{\bar{q}}\rangle}{m_q
m_{\bar{q}}}, \label{swave}\\
 M_{L\not =0}&=&m_q+m_{\bar{q}}+a_L+b_L\left({1\over m_q}+{1\over
m_{\bar{q}}}\right)+
 c_L\left({1\over m^2_q}+{1\over m^2_{\bar{q}}}\right)+
 {d_L\over{m_qm_{\bar{q}}}}+e_L\frac{\langle\mbox{\bf s}_q\cdot\mbox{\bf s}_{\bar{q}}\rangle}{m_q
m_{\bar{q}}}\nonumber\\ &+& f_L\left(\frac{1}{m^3_q}+\frac{1}{m^3_{\bar{q}}}\right)
+g_L\left[\frac{(m_q+m_{\bar{q}})^2+2m_qm_{\bar{q}}}{4m^2_qm^2_{\bar{q}}}\langle\mbox{\bf
L}\cdot\mbox{\bf S}\rangle -\frac{m^2_q-m^2_{\bar{q}}}{4m^2_qm^2_{\bar{q}}}\langle\mbox{\bf
L}\cdot\mbox{\bf S}_-\rangle\right ]\nonumber\\
&+&\frac{h_L}{m_qm_{\bar{q}}}\langle\mbox{\bf S}_{q\bar{q}}\rangle, \label{pwave}
\end{eqnarray}
 where $m_q$ and $m_{\bar{q}}$ are the constituent quark masses, $\mbox{\bf s}_q$ and
 $\mbox{\bf s}_{\bar{q}}$ are the constituent quark spins,
 $e_0$, $a_L$, $b_L$, $c_L$, $d_L$, $e_L$, $f_L$, $g_L$ and $h_L$ are constants,
 $\mbox{\bf S}=\mbox{\bf s}_q+\mbox{\bf s}_{\bar{q}}$, $\mbox{\bf S}_-=\mbox{\bf s}_q-\mbox{\bf
 s}_{\bar{q}}$, and
$\mbox{\bf S}_{q\bar{q}}=3\frac{(\mbox{\bf s}_q\cdot\mbox{\bf r})(\mbox{\bf
s}_{\bar{q}}\cdot\mbox{\bf r})}{r^2}-\mbox{\bf s}_q\cdot\mbox{\bf s}_{\bar{q}}$. Angular momentum
part of the matrix elements of (\ref{swave}) and (\ref{pwave}) is shown in Table 1.
\begin{table}[htb]
\begin{center}
\begin{tabular}{c|cccccc}\hline
         & $^3P_2$& $^3P_1$ &$^3P_0$& $^1P_1$&$^3S_1$&$^1S_0$\\\hline
$\langle\mbox{\bf s}_q\cdot\mbox{\bf s}_{\bar{q}}\rangle$&$\frac{1}{4}$&$\frac{1}{4}$&$\frac{1}{4}$&$-\frac{3}{4}$&$\frac{1}{4}$&$-\frac{3}{4}$\\
  $\langle\mbox{\bf L}\cdot\mbox{\bf S}\rangle$&1&$-1$&$-2$&0\\
 $\langle\mbox{\bf S}_{q\bar{q}}\rangle$&$-\frac{2}{5}$&2&$-4$&0\\
$\langle\mbox{\bf L}\cdot\mbox{\bf S}_-\rangle$&0&0&0&$\frac{3}{2}$\\
 \hline
\end{tabular}
\caption{\small Angular momentum part of the matrix elements of (\ref{swave}) and (\ref{pwave}).}
\end{center}
\end{table}

 With the help of Table 1,
applying (\ref{swave}) and (\ref{pwave}) to $S$-wave and $P$-wave mesons, in the SU(2) flavor
symmetry limit, one can obtain\footnote{Where $n\bar{n}=(u\bar{u}+d\bar{d})/\sqrt{2}$, All the
masses used as input in the present work are taken from PDG\cite{pdg}.}
\begin{eqnarray}
\frac{M_{\pi}+3M_{\rho}}{2M_K+6M_{K^\ast}-M_{\pi}-3M_{\rho}}=\frac{m_u}{m_s}=0.6298\pm 0.00068,
\end{eqnarray}
and
\begin{eqnarray}
\frac{M(^3P_2)_{s\bar{s}}-M(^1P_1)_{s\bar{s}}}{M(^3P_2)_{n\bar{n}}-M(^1P_1)_{n\bar{n}}}=\frac{m_u^2}{m^2_s}.
\label{mass-1}
\end{eqnarray}
From (\ref{mass-1}), with the help of the Gell-Mann-Okubo mass formula\cite{okubo}
\begin{eqnarray}
&&M^2(^3P_2)_{s\bar{s}}+M^2(^3P_2)_{n\bar{n}}=2M^2_{K(^3P_2)},\\
&&M^2(^1P_1)_{s\bar{s}}+M^2(^1P_1)_{n\bar{n}}=2M^2_{K_1(^1P_1)},
\end{eqnarray}
taking $M(^3P_2)_{n\bar{n}}=M_{a_2(1320)}=1318.3\pm 0.6$ MeV,
$M(^1P_1)_{n\bar{n}}=M_{b_1(1235)}=1229.5\pm 3.2$ MeV and $M_{K(^3P_2)}=M_{K^\ast_2(1430)}=1429\pm
0.99$ MeV , one can arrive that
\begin{eqnarray}
M_{K(^1P_1)}=1369.52\pm 1.92~ \mbox{MeV}. \label{m1p1ns}
\end{eqnarray}

The $K_1(^3P_1)$ and $K_1(^1P_1)$ can mix to produce the physical states $K_1(1400)$ and
$K_1(1270)$ and the mixing between $K_1(^3P_1)$ and $K_1(^1P_1)$ can be parameterized
as\cite{suzuki}
\begin{eqnarray}
\begin{array}{ll}
K_1(1400)=&K_1(^3P_1)\cos\theta_K-K_1(^1P_1)\sin\theta_K,\\
K_1(1270)=&K_1(^3P_1)\sin\theta_K+K_1(^1P_1)\cos\theta_K, \label{mixing}
\end{array}
\end{eqnarray}
where $\theta_K$ denotes the $K_1(^3P_1)-K_1(^1P_1)$ mixing angle.  Without any assumption about
the origin of the $K_1(^3P_1)-K_1(^1P_1)$ mixing,
 the masses
of the $K_1(^3P_1)$ and $K_1(^1P_1)$ can be related to $M_{K_1(1400)}$ and $M_{K_1(1270)}$, the
masses of the $K_1(1400)$ and $K_1(1270)$, by the following relation phenomenologically,
\begin{eqnarray}
S \left(\begin{array}{cc}
M^2_{K_1(^3P_1)}& A\\
A&M^2_{K_1(^1P_1)}
\end{array}\right)
S^\dagger= \left(\begin{array}{cc}
M^2_{K_1(1400)}&0\\
0&M^2_{K_1(1270)}
\end{array}\right),
\label{kakb}
\end{eqnarray}
where $A$ denotes a parameter describing the $K_1(^3P_1)-K_1(^1P_1)$ mixing , and
\begin{eqnarray}
S=\left(\begin{array}{cc}
\cos\theta_K&-\sin\theta_K\\
\sin\theta_K&\cos\theta_K
\end{array}\right).
\nonumber
\end{eqnarray}
From (\ref{kakb}), one can have
\begin{eqnarray}
&&M^2_{K_1(^3P_1)}=M^2_{K_1(1400)}\cos^2\theta_K+M^2_{K_1(1270)}\sin^2\theta_K,
\label{kmass1}\\
&&M^2_{K_1(^1P_1)}=M^2_{K_1(1400)}\sin^2\theta_K+M^2_{K_1(1270)}\cos^2\theta_K,
\label{kmass2}\\
&&\cos(2\theta_K)=\frac{M^2_{K_1(^3P_1)}-M^2_{K_1(^1P_1)}}{M^2_{K_1(1400)}-M^2_{K_1(1270)}}.
\label{kmass3}
\end{eqnarray}
Inputting $M_{K_1(1400)}=1402\pm 7$ MeV, $M_{K_1(1270)}=1273\pm 7$ MeV and $M_{K_1(^3P_1)}\simeq
1369.52\pm 1.92$ MeV shown in (\ref{m1p1ns}), from (\ref{kmass1})-(\ref{kmass3}), we have
\begin{eqnarray}
M_{K_1(^3P_1)}= 1307.88\pm 10.33~\mbox{MeV},~~\theta_K=\pm (59.29\pm 2.87)^\circ. \label{k1p1}
\label{mixangle}
\end{eqnarray}

Obviously, the present result that $(M_{K_1(^1P_1)}, M_{K_1(^3P_1)})=(1369.5\pm 1.92, 1307.88\pm
10.33)$ MeV and $\theta_K=\pm (59.29\pm 2.87)^\circ$ is in good agreement with that
$(M_{K_1(^1P_1)}, M_{K_1(^3P_1)})=(1370.03\pm 9.69, 1307.35\pm 0.63)$ MeV and $\theta_K=\pm
(59.55\pm 2.81)^\circ$ given by Ref.\cite{epja26} based on the $f_1(1285)-f_1(1420)$ mixing angle
$\sim 50^\circ$ extracted from the analysis for a substantial body of data concerning the
$f_1(1420)$ and $f_1(1285)$\cite{zphysc76}.

Within the nonrelativistic constituent quark model, the results regarding the masses of the
$K_1(^1P_1)$ and $K_1(^3P_1)$,  $(M_{K_1(^1P_1)}, M_{K_1(^3P_1)})=(1368, 1306)$ MeV suggested by
\cite{nrqm} and $(M_{K_1(^1P_1)}, M_{K_1(^3P_1)})=(1356, 1322)$ MeV suggested by \cite{prd57}, are
in good agreement with our predicted result. However, based on the following relation employed by
\cite{nrqm,prd57}
\begin{eqnarray}
\tan^2(2\theta_K)=\left(\frac{M^2_{K_1(^3P_1)}-M^2_{K_1(^1P_1)}}{M^2_{K_1(1400)}-M^2_{K_1(1270)}}\right)^2-1,
\label{tan}
\end{eqnarray}
the values of $\theta_K = (31\pm 4)^\circ$ given by\cite{nrqm} and $\theta_K=(37.3\pm 3.2)^\circ$
given by\cite{prd57} disagree with value of $|\theta_K|\simeq (59.29\pm 2.87) ^\circ$ given by the
present work.

As pointed out by our previous paper \cite{epja26}, (\ref{tan}) is equivalent to (\ref{kmass3}),
and will yield two solutions $|\theta_K|$ and $\frac{\pi}{2}-|\theta_K|$. Simultaneously
considering the relations (\ref {kmass1}), (\ref{kmass2}) and (\ref{tan}), in the presence of
$M_{K_1(1400)}>M_{K_1(1270)}$, we can conclude that if $M_{K_1(^3P_1)}< M_{K_1(^1P_1)}$, the
$|\theta_K|$ would greater than $45^\circ$ . In fact, relation (\ref{kmass3}) clearly indicates
that in the presence of $M_{K_1(1400)}>M_{K_1(1270)}$, the case $M_{K_1(^3P_1)}< M_{K_1(^1P_1)}$
must require $45^\circ <|\theta_K|< 90^\circ$.

\section{The sign of $\theta_K$ constrained by experimental information}
\indent\vspace*{-1cm}

 Now we wish to discuss the sign of $\theta_K$ by considering the open-flavor strong
decays of the $K_1(1270)$ and $K_1(1400)$ in the $^3P_0$ decay model, and the production ratio of
these two physical strange states in the $\tau$ decay.
\subsection{Strong decays of the $K_1(1270)$ and $K_1(1400)$ in the $^3P_0$ model}
\indent\vspace*{-1cm}

The main assumption of the $^3P_0$ decay model is that the strong decays
 take place via the production of a quark-antiquark pair with the vacuum quantum numbers which corresponds to the $^3P_0$ state of a quark-antiquark pair.
 After the $^3P_0$ decay model was originally introduced by Micu\cite{Micu},
 it was applied
extensively to meson and baryon decays. It is widely accepted that the $^3P_0$ model is successful
since it gives a good description of many of the observed decay amplitudes and partial widths of
the open-flavor meson strong decays.

Assuming a fixed $^3P_0$ source strength, simple harmonic oscillator quark model meson wave
functions and physical phase space, Ackleh et al.\cite{t1} developed a diagrammatic, momentum-space
formulation of the $^3P_0$ model to evaluate the partial width $\Gamma_{A\rightarrow BC}$
\begin{eqnarray}
 \Gamma_{A\rightarrow BC}=2\pi\frac{PE_BE_C}{M_A}\sum\limits_{LS}|{\cal{M}}_{LS}|^2,
\label{width}
\end{eqnarray}
where $P$ is the decay momentum for the decay $A\rightarrow B+C$, $E_B$ and $E_C$ are the energies
of mesons $B$ and $C$, in the rest frame of $A$,
\begin{eqnarray}
&&P=\frac{[(M^2_A-(M_B+M_C)^2)(M^2_A-(M_B-M_C)^2)]^{1/2}}{2M_A},\nonumber\\
&&E_B=(M^2_A-M^2_C+M^2_B)/2M_A,\nonumber\\
&&E_C=(M^2_A-M^2_B+M^2_C)/2M_A,\nonumber
\end{eqnarray}
$M_A$, $M_B$ and $M_C$ denote the masses of the mesons $A$, $B$ and $C$, respectively;
 ${\cal{M}}_{LS}$ are
proportional to an overall Gaussian in $x=P/\beta$ times a channel-dependent polynomial
${\cal{P}}_{LS}$,
\begin{eqnarray}
{\cal{M}}_{LS}=\frac{\gamma}{\pi^{1/4}\beta^{1/2}}{\cal{P}}_{LS}(x)e^{-x^2/12}.\nonumber
\end{eqnarray}
 It is found that this formulation with the width parameter $\beta=0.4$ GeV and the
pair-production strength parameter $\gamma=0.4$ can give a reasonably accurate description of the
overall scale of decay widths\cite{t2,t3}.

Based on (\ref{mixing}) and (\ref{width}), employing the analytical results for ${\cal{P}}_{LS}$
listed in Appendix A of Ref.\cite{t2}, one can have\cite{t3}
\begin{eqnarray}
\Gamma(K_1(1270)\rightarrow \rho
K)&=&21.8\cos^2\theta_K+61.6\sin\theta_K\cos\theta_K+43.6\sin^2\theta_K,\\
\Gamma(K_1(1270)\rightarrow \pi
K^\ast)&=&59.6\cos^2\theta_K-158.7\sin\theta_K\cos\theta_K+115.7\sin^2\theta_K,\\
\Gamma_{\mbox{thy}}(K_1(1270))&=&81\cos^2\theta_K-97\sin\theta_K\cos\theta_K+159\sin^2\theta_K,\\
\Gamma(K_1(1400)\rightarrow \rho
K)&=&160\cos^2\theta_K-219.9\sin\theta_K\cos\theta_K+82.3\sin^2\theta_K,\\
\Gamma(K_1(1400)\rightarrow \omega
K)&=&52.3\cos^2\theta_K-72.3\sin\theta_K\cos\theta_K+26.8\sin^2\theta_K,\\
\Gamma(K_1(1400)\rightarrow \pi
K^\ast)&=&141.1\cos^2\theta_K+176.2\sin\theta_K\cos\theta_K+78.8\sin^2\theta_K,\\
\Gamma_{\mbox{thy}}(K_1(1400))&=&353\cos^2\theta_K-116\sin\theta_K\cos\theta_K+188\sin^2\theta_K,\\
|D/S|^2&=&\left\{\begin{array}{ll}
\frac{(-0.0411\cos\theta_K-0.029\sin\theta_K)^2}{(-0.204\cos\theta_K+0.288\sin\theta_K)^2},&\mbox{for}~K_1(1270)\rightarrow
\pi K^\ast\\
\frac{(-0.0498\cos\theta_K+0.0704\sin\theta_K)^2}{(+0.247\cos\theta_K+0.175\sin\theta_K)^2},&\mbox{for}~K_1(1400)\rightarrow
\pi K^\ast
\end{array}\right..
\end{eqnarray}

 For $\theta_K=\pm(59.29\pm 2.87)^\circ$, the theoretical results regarding the above widths
are shown in Tables 2, 3 and 4. Tables 2-4 clearly indicate that the present experimental data
strongly prefer $\theta_K=+(59.29\pm 2.87)^\circ$ over $\theta_K=-(59.29\pm 2.87)^\circ$.
\begin{table}[hbt]
\begin{center}
\begin{tabular}{llll}\hline
   $K_1(1270)$ & Exp.\cite{pdg}  & $\theta_K=+(59.29\pm 2.87)^\circ$        &$\theta_K=-(59.29\pm 2.87)^\circ$\\\hline
  $\Gamma$ (MeV)              &{\bf 90}$\pm${\bf 20}   &  $96.07\pm5.76$      &$181.25\pm 1.11$\\
  $\Gamma(\rho K)/\Gamma(\pi K^\ast)$     &{\bf 2.625}$\pm${\bf 0.902}& $2.07\pm 0.41$ &$0.064\pm 0.014$\\
  \hline
\end{tabular}
\caption{\small The predicted results of the $K_1(1270)$ strong decays in the $^3P_0$ decay model.}
\end{center}
\end{table}
\begin{table}[hbt]
\begin{center}
\begin{tabular}{llll}\hline
   $K_1(1400)$ & Exp.\cite{pdg}  & $\theta_K=+(59.29\pm 2.87)^\circ$        &$\theta_K=-(59.29\pm 2.87)^\circ$\\\hline
  $\Gamma$ (MeV)              &{\bf 174}$\pm${\bf 13}   &  $180.1\pm 4.48$      &$282.0\pm 10.0$\\
  $\Gamma(\rho K)/\Gamma$     &{\bf 0.03}$\pm${\bf 0.03}& $0.033\pm 0.01$      &$0.71\pm 0.04$\\
  $\Gamma(\omega K)/\Gamma$     &{\bf 0.01}$\pm${\bf 0.01}& $0.0095\pm 0.0034$  &$0.23\pm 0.01$\\
  $\Gamma(\pi K^\ast)/\Gamma$     &{\bf 0.94}$\pm${\bf 0.06}& $0.96\pm 0.05$   &$0.063\pm 0.006$\\
  \hline
\end{tabular}
\caption{\small The predicted results of the $K_1(1400)$ strong decays in the $^3P_0$ decay model.}
\end{center}
\end{table}
\begin{table}[hbt]
\begin{center}
\begin{tabular}{llll}\hline
   $|D/S|^2$ & Exp.\cite{pdg}  & $\theta_K=+(59.29\pm 2.87)^\circ$        &$\theta_K=-(59.29\pm 2.87)^\circ$\\\hline
  $K_1(1270)\rightarrow \pi K^\ast$               &{\bf 1.0}$\pm${\bf 0.7}   &  $0.1\pm 0.03$      &$0.0001\pm 0.0002$\\
  $K_1(1400)\rightarrow \pi K^\ast$               &{\bf 0.04}$\pm${\bf 0.01}   &  $0.02\pm 0.004$      &$12.5\pm 15.6$\\
  \hline
\end{tabular}
\caption{\small The $|D/S|^2$ ratios for $K_1(1270)\rightarrow \pi K^\ast$ and
$K_1(1400)\rightarrow \pi K^\ast$ in the $^3P_0$  model.}
\end{center}
\end{table}

\subsection{Production ratio of the $K_1(1270)$ and $K_1(1400)$ in the $\tau$ decay}
 \indent\vspace*{-1cm}

With the definition of the decay constant of the axial-vector meson given by\cite{cheng}
\begin{eqnarray}
\langle 0|A_\mu|A(q,\varepsilon)\rangle=f_A m_A\varepsilon_\mu,
\end{eqnarray}
the partial width for $\tau\rightarrow \nu_\tau K_1$ can be expressed by
\begin{eqnarray}
\Gamma(\tau\rightarrow\nu_\tau K_1)=\frac{G^2_F}{16\pi}|V_{us}|^2 f^2_{K_1}
\frac{(m^2_\tau+2m^2_{K_1})(m^2_\tau-m^2_{K_1})^2}{m^3_\tau}. \label{twidth}
\end{eqnarray}
 Considering the
SU(3) breaking corrections, following Ref.\cite{cheng,suzuki}, we have
\begin{eqnarray}
\frac{f_{K_1(1270)}m_{K_1(1270)}}{f_{K_1(1400)}m_{K_1(1400)}}=\frac{\sin\theta_K-\delta\cos\theta_K}{\cos\theta_K+\delta\sin\theta_K},
\label{breaking}
 \end{eqnarray}
 where the parameter $\delta$ denoting a SU(3) breaking factor has the following form in the static limit of the quark model\cite{blundell}\footnote{$m_u=307.8\pm 0.19$ MeV and $m_s=488.69\pm 0.28$ MeV derived from
(\ref{swave}).}
\begin{eqnarray}
\delta=\frac{1}{\sqrt{2}}\frac{m_s-m_u}{m_s+m_u}=0.16\pm 0.0003.
\end{eqnarray}
From (\ref{twidth}) and (\ref{breaking}),
 the $K_1(1400)$ and $K_1(1270)$ production ratio in the
$\tau$ decay can be given by
\begin{eqnarray}
\frac{\Gamma(\tau\rightarrow \nu_\tau K_1(1270))}{\Gamma(\tau\rightarrow \nu_\tau K_1(1400))}=
F_p\left|\frac{\sin\theta_K-\delta\cos\theta_K}{\cos\theta_K+\delta\sin\theta_K}\right |^2,
\label{taudecay}
\end{eqnarray}
where $F_p$ denotes the phase factor given by
\begin{eqnarray}
F_p=\frac{\left(m^2_\tau+2m^2_{K_1(1270)}\right)\left(m^2_\tau-m^2_{K_1(1270)}\right)^2m^2_{K_1(1400)}}{\left(m^2_\tau+2m^2_{K_1(1400)}\right)\left(m^2_\tau-m^2_{K_1(1400)}\right)^2m^2_{K_1(1270)}}=1.82\pm
0.086.\nonumber
\end{eqnarray}
Then, from (\ref{taudecay}), one can have
\begin{eqnarray}
\frac{\Gamma(\tau\rightarrow \nu_\tau K_1(1270))}{\Gamma(\tau\rightarrow \nu_\tau
K_1(1400))}=\left\{
\begin{array}{rrr}
2.62\pm 0.55,&\mbox{for}& \theta_K=+(59.29\pm 2.87)^\circ\\
11.59\pm 3.43,&\mbox{ for}& \theta_K=-(59.29\pm 2.87)^\circ
\end{array}\right..
\label{tautheory}
\end{eqnarray}

Experimentally, the ${\cal{B}}(\tau\rightarrow\nu_\tau K_1(1270))$ and
${\cal{B}}(\tau\rightarrow\nu_\tau K_1(1400))$ have been reported by TPC/Two-Gamma
collaboration\cite{TPC} in 1994 and ALEPH collaboration\cite{ALEPH} in 1999, respectively. The
averaged result of these two collaborations is given by\cite{pdg}
\begin{eqnarray}
\left.\begin{array}{c}
{\cal{B}}(\tau\rightarrow\nu_\tau K_1(1270))=(0.47\pm 0.11)\times 10^{-2}\\
{\cal{B}}(\tau\rightarrow\nu_\tau K_1(1400))=(0.17\pm 0.26)\times 10^{-2}
\end{array}\right.,
\end{eqnarray}
which gives that
\begin{eqnarray}
\left.\frac{\Gamma(\tau\rightarrow \nu_\tau K_1(1270))}{\Gamma(\tau\rightarrow \nu_\tau
K_1(1400))}\right|_{\mbox{exp-1}}=2.76\pm 4.28. \label{tauexp-1}
\end{eqnarray}
This measured result also is in favor of $\theta_K=+(59.29\pm 2.87)^\circ$ over
 $\theta_K=-(59.29\pm 2.87)^\circ$, although the uncertainty of the reported result is large as shown in (\ref{tauexp-1}).

 Assuming the resonance structure of $\tau^-\rightarrow K^-\pi^+\pi^-\nu_\tau$
 decays being  dominated by the $K_1(1270)$ and $K_1(1400)$ resonances, in 2000, both CLEO collaboration\cite{asner} and OPAL
  collaboration\cite{OPAL} have also measured the ratio of $\nu_\tau K_1(1270)$ to
  $\nu_\tau K_1(1400)$ with the averaged result\cite{pdg}
\begin{eqnarray}
\frac{\Gamma(\tau\rightarrow \nu_\tau K_1(1270))}{\Gamma(\tau\rightarrow \nu_\tau
K_1(1270))+\Gamma(\tau\rightarrow \nu_\tau K_1(1400))}=0.69\pm 0.15,
\end{eqnarray}
which therefore in turn implies that
\begin{eqnarray}
\left.\frac{\Gamma(\tau\rightarrow \nu_\tau K_1(1270))}{\Gamma(\tau\rightarrow \nu_\tau
K_1(1400))}\right|_{\mbox{exp-2}}=2.23\pm 1.56. \label{tauexp}
\end{eqnarray}
Comparison of (\ref{tautheory}) and (\ref{tauexp}) again shows that the present experimental data
strongly prefer $\theta_K=+(59.29\pm 2.87)^\circ$ over $\theta_K=-(59.29\pm 2.87)^\circ$.

It is based on the $K_1(1400)$ production dominance in the $\tau$ decay that Suzuki suggests that
the preferred result is $\theta_K\approx 33^\circ$ rather than $57^\circ$\cite{suzuki}. However,
the recent available experiment data shown in (\ref{tauexp}) clearly show the $K_1(1270)$ dominance
in the $\tau$ decay. Consequently, the argument of ruling out $\theta_K\approx 57^\circ$ from the
$K_1(1400)$ dominance is therefore no longer valid. The study of hadronic decays $D\rightarrow
K_1(1270)\pi, K_1(1400)\pi$ decays performed by Cheng\cite{cheng} favors $\theta_K\approx
-58^\circ$, however as pointed out by Cheng et al. in Ref.\cite{cheng1} that this argument is
subject to many uncertainties such as the unknown $D\rightarrow K_1(^1P_1),K_1(^3P_1)$ transition
form factors and the decay constants of $K_1(1270)$ and $K_1(1400)$. We note that the recent
analysis for the SU(3) nonets of the axial vector mesons into a vector and a pseudoscalar performed
by Roca et al.\cite{Roca} based on a tensor formulation of the vector and axial vector fields gives
$\theta_K=+(62\pm 3)^\circ$, which is in fact in good agreement with our suggested result that
$\theta_K=+(59.29\pm 2.87)^\circ$.

\section{Concluding remarks}
\indent \vspace*{-1cm}

In the nonrelativistic constituent quark model, the masses of the $K_1(^3P_1)$ and $K_1(^1P_1)$ are
determined to be $1307.88\pm 10.33$ and $1396.5\pm 1.92$ MeV, respectively, which therefore
suggests that the absolute value of the $K_1(^3P_1)-K_1(^1P_1)$ mixing angle is $(59.29\pm
2.87)^\circ$. These findings are in good agreement with those given by Ref.\cite{epja26} based on
the investigation on the implication of the $f_1(1285)-f_1(1420)$ mixing for the
$K_1(^3P_1)-K_1(^1P_1)$ mixing angle. Investigating the open-flavor strong decays of the
$K_1(1270)$ and $K_1(1400)$ in the $^3P_0$ decay model, we find the current experimental data
strongly prefer $\theta_K=+(59.29\pm2.87)^\circ$ over $\theta_K=-(59.29\pm2.87)^\circ$. The
analysis for the production ratio of the $K_1(1270)$ and $K_1(1400)$ in the $\tau$ decay also
indicates that the experimental data is in favor of the result $\theta_K=+(59.29\pm2.87)^\circ$.

In the framework of a covariant light-front quark model, the calculations performed by Cheng et
al.\cite{cheng1} for the exclusive radiative $B$ decays, $B\rightarrow K_1(1270)\gamma$,
$K_1(1400)\gamma$, show that the relative strength of $B\rightarrow K_1(1270)\gamma$ and
$B\rightarrow K_1(1270)\gamma$ rates is very sensitive to the sign of the $\theta_K$. The recent
analysis of two-body $B$ decays with an axial-vector meson in the final state performed by Nardulli
et al.\cite{Nard,0602243} using naive factorization, shows the branching ratios for $B\rightarrow
b_1\pi$, $b_1K$, $a_1\pi$ and $a_1K$ also depend strongly on the $\theta_K$. In addition, as
pointed by Suzuki\cite{suzuki1}, the relation $|Am(J/\psi(\psi^\prime)\rightarrow$
$K^0_1(1400)\overline{K^0}|^2=\tan^2\theta_K|Am(J/\psi(\psi^\prime)\rightarrow
K^0_1(1270)\overline{K^0}|^2$ can be able to determine the $\theta_K$ directly without referring to
other parameters. Therefore, in order to further check the consistency of our suggested mixing
angle of $K_1(1270)$ and $K_1(1400)$, detailed experimental study of the above mentioned decays
involving the axial-vector mesons is certainly desirable.

 \noindent {\bf Acknowledgments:}
This work is supported in part by National Natural Science Foundation of China under Contract No.
10205012, Henan Provincial Science Foundation for Outstanding Young Scholar under Contract No.
0412000300, Henan Provincial Natural Science Foundation under Contract No. 0311010800, Foundation
of the Education Department of Henan Province under Contract No. 2003140025, and the Program for
Youthful Excellent Teachers in University of Henan Province.

\end{document}